\documentclass[sigconf]{acmart}

\usepackage{amsfonts}
\usepackage{algorithmic}
\usepackage{textcomp}
\usepackage{enumitem}
\usepackage[table]{xcolor}
\usepackage[ruled,linesnumbered]{algorithm2e}
\usepackage{makecell}

\usepackage{subcaption}
\usepackage{xspace}
\usepackage{multirow}
\usepackage{url}
\usepackage{cleveref}
\usepackage{listings}
\usepackage{fontawesome5}
\hypersetup{hidelinks,
	colorlinks=true,
	allcolors=black,
	pdfstartview=Fit,
	breaklinks=true,
 urlcolor=blue}

\definecolor{codeblue}{rgb}{0.1, 0.1, 0.6}
\definecolor{codecomment}{gray}{0.45}
\definecolor{placeholder}{rgb}{0.8, 0.4, 0.1}

\definecolor{codebg}{RGB}{248,248,248}   
\definecolor{codeframe}{RGB}{200,200,200} 
\definecolor{keywordcolor}{RGB}{0,0,160}  
\definecolor{commentcolor}{RGB}{0,120,0}  
\definecolor{stringcolor}{RGB}{163,21,21} 

\lstdefinestyle{custom}{
    breakable,
    backgroundcolor=\color{codebg},
    frame=single,
    rulecolor=\color{codeframe},
    basicstyle=\ttfamily\footnotesize,
    keywordstyle=\color{keywordcolor}\bfseries,
    commentstyle=\color{commentcolor}\itshape,
    stringstyle=\color{stringcolor},
    numberstyle=\tiny\color{gray},
    numbers=left,
    numbersep=5pt,
    tabsize=4,
    breaklines=true,
    showstringspaces=false,
    captionpos=b
}

\hypersetup{colorlinks = true, 
	    linkcolor = blue, 
	    urlcolor = blue,
            citecolor = red} 
\restylefloat{figure}
\AtBeginDocument{%
  }

\copyrightyear{2026}
\acmYear{2026}
\setcopyright{cc}
\setcctype{by}
\acmConference[ICSE-SEIP '26]{2026 IEEE/ACM 48th International Conference on Software Engineering}{April 12--18, 2026}{Rio de Janeiro, Brazil}
\acmBooktitle{2026 IEEE/ACM 48th International Conference on Software Engineering (ICSE-SEIP '26), April 12--18, 2026, Rio de Janeiro, Brazil}
\acmPrice{}
\acmDOI{10.1145/3786583.3786856}
\acmISBN{979-8-4007-2426-8/2026/04}

\begin{document}
\title{\textsc{EvoC2Rust}: A Skeleton-guided Framework for Project-Level C-to-Rust Translation}

\author{Chaofan Wang}
\affiliation{%
  \institution{Shanghai Jiao Tong University}
  \city{Shanghai}
  \country{China}}
\email{chaofwang@sjtu.edu.cn}

\author{Tingrui Yu}
\affiliation{%
  \institution{Shanghai Jiao Tong University}
  \city{Shanghai}
  \country{China}}
\email{hzfsls@sjtu.edu.cn}

\author{Beijun Shen}\authornote{Corresponding author.}
\affiliation{%
  \institution{Shanghai Jiao Tong University}
  \city{Shanghai}
  \country{China}}
\email{bjshen@sjtu.edu.cn}

\author{Jie Wang}
\affiliation{%
  \institution{Huawei Technologies Co., Ltd}
  \city{Beijing}
  \country{China}}
\email{wangjie451@huawei.com}

\author{Dong Chen}
\affiliation{%
  \institution{Huawei Technologies Co., Ltd}
  \city{Beijing}
  \country{China}}
\email{chendong108@huawei.com}

\author{Wenrui Zhang}
\affiliation{%
  \institution{Huawei Technologies Co., Ltd}
  \city{Beijing}
  \country{China}}
\email{zhangwenrui8@huawei.com}

\author{Yuling Shi}
\affiliation{%
  \institution{Shanghai Jiao Tong University}
  \city{Shanghai}
  \country{China}}
\email{yuling.shi@sjtu.edu.cn}

\author{Chen Xie}
\affiliation{%
  \institution{Shanghai Jiao Tong University}
  \city{Shanghai}
  \country{China}}
\email{dindin\_xc@sjtu.edu.cn}

\author{Xiaodong Gu}
\affiliation{%
  \institution{Shanghai Jiao Tong University}
  \city{Shanghai}
  \country{China}}
\email{xiaodong.gu@sjtu.edu.cn}

\renewcommand{\shortauthors}{Wang et al.}

\begin{abstract}
Translating legacy C codebases to Rust is increasingly demanded for building safety-critical systems. While various approaches have emerged for this task, they face inherent trade-offs: rule-based methods often struggle to satisfy code safety and idiomaticity requirements, while LLM-based methods frequently fail to generate semantically equivalent Rust code, due to the heavy dependencies of modules across the entire codebase. Recent studies have revealed that both solutions are limited to small-scale programs.
In this paper, we propose \textsc{EvoC2Rust}, an automated framework for converting complete C projects to equivalent Rust ones. \textsc{EvoC2Rust} employs a skeleton-guided translation strategy for project-level translation. The pipeline consists of three stages: 1) it first decomposes the C project into functional modules, employs a feature-mapping-enhanced LLM to transform definitions and macros, and generates type-checked function stubs, which form a compilable Rust skeleton; 2) it then incrementally translates functions, replacing the corresponding stub placeholders; 3) finally, it repairs compilation errors by integrating LLM and static analysis.
Through evolutionary augmentation, \textsc{EvoC2Rust} combines the advantages of both rule-based and LLM-based solutions. 
Our evaluation on open-source benchmarks and six industrial projects demonstrates the superior performance of \textsc{EvoC2Rust} in project-level C-to-Rust translation. 
The results show that our approach outperforms the strongest LLM-based baseline by 17.24\% in syntax accuracy and 14.32\% in semantic accuracy, while also achieving a 43.59\% higher code safety rate than the best rule-based tool.
\end{abstract}



\maketitle

\section{Introduction}

The translation of legacy C code to Rust is gaining significant attention in software engineering, spurred by critical memory safety concerns~\cite{LiW25, DARPA}. Such vulnerabilities constitute the most prevalent category of critical security defects in production C systems, accounting for approximately 70\% of high-severity vulnerabilities in industry reports from Google and Microsoft~\cite{UrgentNeed}. This has accelerated a paradigm shift toward compile-time safety guarantees through safe programming languages~\cite{WhiteHouse}. In response, Rust has emerged as a leading alternative that offers memory safety by enforcing a strict ownership and borrowing model~\cite{Williams24a}.

\begin{figure*}[ht]
\centerline{\includegraphics[width=0.95\textwidth, trim=0 0 0 0, clip]{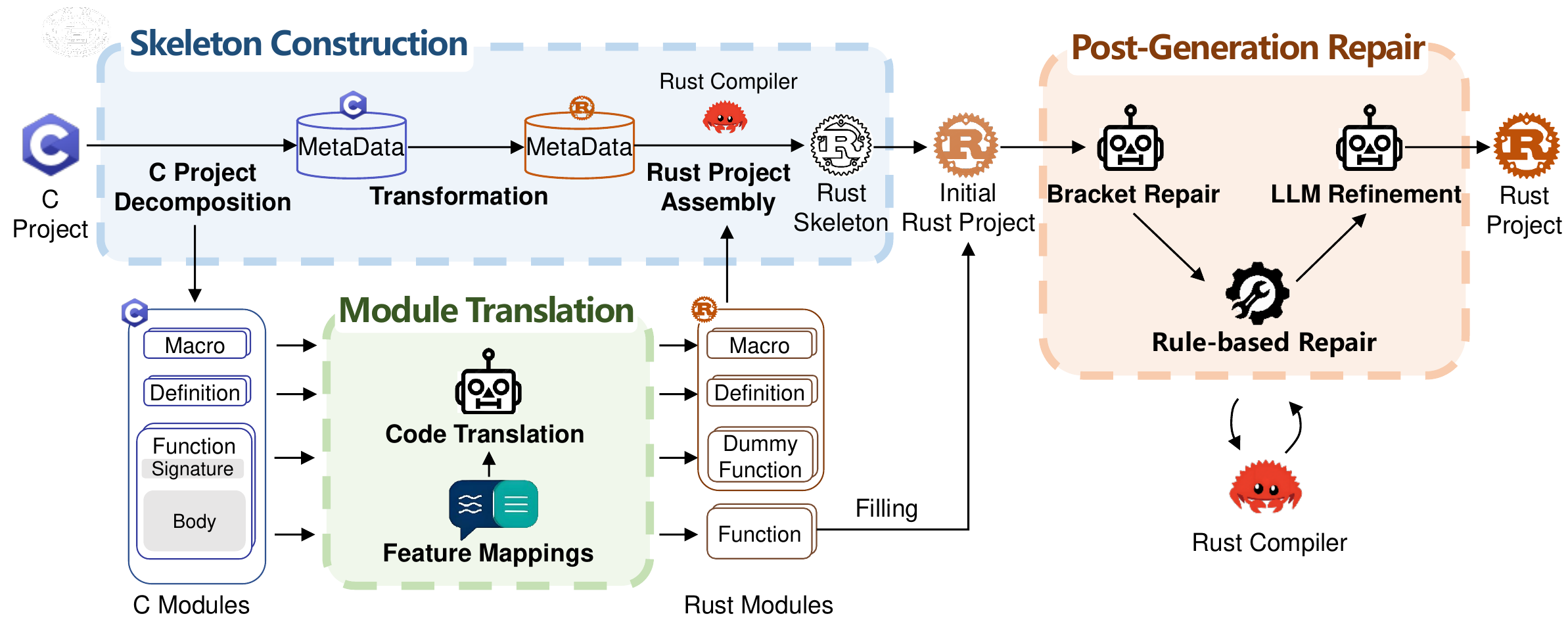}}
    \caption{Overview of \textsc{EvoC2Rust}.}
    \label{fig:overview}
    \Description{The figure illustrates the workflow of EvoC2Rust. First, a C project is decomposed into modules containing macros, definitions, function signatures, and bodies. The extracted metadata is transformed and assembled into an initial Rust skeleton. Concurrently, module-level code translation employs feature mappings to generate corresponding Rust modules, which are integrated into the skeleton to form an initial Rust project. This project then undergoes a post-generation repair stage consisting of bracket repair, rule-based repair, and LLM refinement, ultimately yielding the final Rust project.}
\end{figure*}

Recent work proposes rule-based tools~\cite{c2rust, ZhangDYW23, HongR23}, which convert C code to Rust using primarily syntax-based transformations on a terse intermediate representation of compilation. Such translation inherently generates non-idiomatic Rust code with low-level constructs and may also contain semantic inaccuracies.  
\cite{LostinTranslation}. 
Another line of work proposes LLM-based techniques~\cite{FLOURINE, LiYGCS24, vert, SpecTra} to utilize the code comprehension capability of code LLMs. Due to the lack of parallel C-Rust training data and significant linguistic differences, these approaches often fail to produce semantically equivalent Rust code~\cite{RustRepoTrans}.

While automated code translation has been extensively investigated~\cite{JiaoYLQGS23, RoziereLCL20, Zhu0R22, Yang0YK0LHMJ024, TaoYGS24, dehghan2025translating}, the translation of C-to-Rust projects poses special challenges:

\textbf{Challenge-1: Substantial linguistic discrepancies for ensuring safety.}
C and Rust differ fundamentally in their safety models. C permits loose type checking, unrestricted pointer arithmetic, and manual memory management, whereas Rust enforces strict compile-time type safety, ownership-based memory guarantees, and disciplined reference usage through its borrowing rules.
Existing rule-based translators often compromise safety by relying on raw pointers, unsafe blocks, or external C functions~\cite{LostinTranslation}.
While LLM-based approaches can generate code with improved safety, they often lack sufficient context understanding (\emph{e.g.}, variable scope or pointer lifetimes), potentially leading to logical errors or memory safety violations in the translated Rust code~\cite{LAC2R}.
As a result, neither rule-based nor LLM-based methods alone reliably meet Rust’s stringent safety requirements.

\textbf{Challenge-2: Project-level code dependence.} 
Translating complete C projects poses greater challenges than translating isolated functions because the translated Rust project must preserve cross-module dependencies and consistent project-level structure~\cite{projectc2rust, Syzygy}.
Code LLMs often struggle with such large-scale context~\cite{YunLGS24}, leading to broken references, inconsistent APIs, or incorrect module boundaries~\cite{LostinTranslation, apirat}.
Consequently, existing approaches tend to perform well only on small-scale programs (\emph{e.g.}, $\le$100 LoC)~\cite{EmreSDH21, FLOURINE}, and fall short in real-world project settings.

In this work, we propose \textsc{EvoC2Rust}, an automated framework for translating complete C projects into equivalent Rust code. 
To address the linguistic gap and ensure safety (Challenges 1), \textsc{EvoC2Rust} enhances LLMs with \emph{safety-preserving mappings} across seven core linguistic features, including \emph{types}, \emph{macros}, \emph{functions}, \emph{operators}, \emph{syntax structures}, \emph{globals}, and \emph{variadic arguments}. 
To enable project-level translation (Challenge 2), \textsc{EvoC2Rust} introduces a skeleton-guided translation strategy that proceeds in three stages. First, it analyzes the input C project to extract high-level metadata and decomposes the project into functional modules, generating a compilable Rust skeleton with placeholder functions.
Next, it incrementally translates these stubs using LLMs augmented with the predefined feature mappings. Finally, it combines LLMs with static analysis to iteratively repair compilation errors and refine the output.
This evolutionary process effectively blends the strengths of rule-based and learning-based approaches, striking a practical balance between correctness and safety in automated C-to-Rust migration.

We evaluate \textsc{EvoC2Rust} using an open-source benchmark Vivo-Bench~\cite{vivo} and a self-collected dataset of six industrial projects. Experimental results demonstrate that \textsc{EvoC2Rust} significantly outperforms existing baselines in project-level C-to-Rust translation. In comparison with LLM-based approaches, \textsc{EvoC2Rust} improves the compilation pass rate by 17.24–50.84\% and the line acceptance rate by 14.32–44.79\%, while also achieving marginally better code safety. When compared to rule-based methods, \textsc{EvoC2Rust} attains a 43.59–96.79\% higher code safety rate, while maintaining competitive performance in terms of compilation and line acceptance rates. Furthermore, at the module level, \textsc{EvoC2Rust} achieves a 92.25\% compilation success rate and an 89.53\% test pass rate on industrial projects.

The contributions of our work are summarized as follows:

\begin{itemize}[leftmargin=20pt, topsep=2pt]
    \item We propose a novel skeleton-guided framework that overcomes the barrier of inter-module dependencies to enable the translation of complete C projects into Rust.

    \item We define safety-preserving mappings between core linguistic features of C and Rust to enhance LLMs' safety awareness. 
    
    \item We evaluate \textsc{EvoC2Rust} on both open-source and industrial scenarios. Experimental results show that \textsc{EvoC2Rust} outperforms existing baselines, effectively translating C projects to correct and safe Rust code.

\end{itemize}

Our source code and experimental data are publicly available at
\url{https://github.com/bbzswcf/EvoC2rust}. 

\begin{table*}[t]
    \centering
    \caption{Type Mappings between C and Rust}
    \begin{tabular}{ll m{4.5cm} l m{6cm}}
    \toprule
     \textbf{Type} & \textbf{C Type} & \textbf{C Example} & \textbf{Rust Type} & \textbf{Rust Example}\\
    \midrule 
    Array & \verb|int[N]| & \verb|int a[3] = {1, 2, 3}| & \verb|Array<i32, N>| & \verb|let mut a: Array<i32,3> = arr![1,2,3]|\\
    Pointer & \verb|int *| & \verb|int *a = &b| & \verb|Ptr<i32>| & \verb|let mut a: Ptr<i32> = c_ref!(b)|\\
    String  & \verb|char *| & \verb|char *c = "Hello World!"| & \verb|Ptr<u8>| & \verb|let mut c=cstr!("Hello World!")|\\
    Function Pointer & \verb|(*func)| & \verb|typedef int (*MyFunc)| \verb|(const void *, const void *)| & \verb|FuncPtr| & \verb|pub type MyFunc=FuncPtr<fn(Ptr<Void>,| \verb|Ptr<Void>) -> i32>|\\
    File & \verb|FILE *| & \verb|FILE *f = fopen(name, "rb")| & \verb|FilePtr| & \verb|let mut f=c_fopen!(name, cstr("rb"))|\\
    Variadic Argument & \verb|va_list| & \verb|va_start(arg_ptr, prev)| & \verb|VaList| & \verb|(…, prev: …, arg_ptr: VaList)|\\
    \bottomrule
    \end{tabular}
    \label{tab:typemapping}
\end{table*}
\section{Approach}
\label{sec:approach}

Figure \ref{fig:overview} illustrates the architecture of \textsc{EvoC2Rust}. It contains three key steps: (1) \emph{Skeleton Construction} decomposes the C project into modules and constructs a Rust project skeleton to assemble the Rust modules (Section~\ref{sec:decomposition}); (2) \emph{Module Translation} incrementally converts definitions, macros, and functions using safety-preserving linguistic mappings (Section~\ref{sec:translation}); and (3) \emph{Post-Generation Repair} combines LLM-based refinement with static analysis for code optimization  (Section~\ref{sec:repair}).

\subsection{Project Skeleton Construction}
\label{sec:decomposition}


A straightforward idea of project-level code translation is to translate each function individually and assemble them into a project. However, naively aggregating translated functions often leads to cascading interdependency errors. To address this, \textsc{EvoC2Rust} employs a skeleton-guided translation strategy, which first constructs a compilable Rust project skeleton based on project metadata and then incrementally populates it with translated function implementations. 

Given a C project, \textsc{EvoC2Rust} first parses the source code using Tree-sitter~\cite{treesitter} to extract key syntactic elements and abstract them into a unified project metadata representation. This metadata captures structural information including filenames, \verb|include| statements, macro definitions, type declarations, global identifiers, function signatures, and a mapping table between identifier declarations to source files. 
They are then systematically transformed into Rust equivalents via rule-based conversion. For instance, C \verb|include| dependencies are transformed to Rust  \verb|use| imports, and identifier declarations are converted into Rust \verb|pub use| statements (\emph{e.g.}, \verb|pub use {Rust_filename}::{identifier_name}|) using the declaration-to-file table.  

Using the converted metadata, \textsc{EvoC2Rust} constructs a Rust project skeleton—comprising Rust constructs automatically translated from C definitions, macros, and function signatures—that accurately preserves the structural organization of the original C project. 
Each C function is assigned a type-checked dummy counterpart in Rust using its translated signature and the \verb|unimplemented!()| macro. These placeholders serve as safe stand-ins, allowing the project to compile successfully even before the actual function bodies are translated.

The skeleton provides a stable framework that allows function bodies to be translated and integrated incrementally. This ensures each addition is contextually correct and free from dependency errors, enabling a seamless and scalable translation process even for large-scale projects with complex interactions.


\subsection{Module Translation with Safety-preserving Mapping}
\label{sec:translation}

\textsc{EvoC2Rust} employs an LLM to automatically translate C modules into Rust constructs, converting definitions, macros, and function signatures to assemble the skeleton, and then transforming function bodies to populate the placeholder dummy functions within the skeleton.
To preserve safety and semantic alignment during translation, \textsc{EvoC2Rust} augments the LLM with a comprehensive set of linguistic mappings across seven key categories: 

1) \textit{Type Mapping}.
Each C type is associated with a semantically consistent Rust counterpart, as summarized in Table~\ref{tab:typemapping}. For example, to represent C-style strings, we introduce a \verb|cstr!()| macro in Rust that guarantees null termination, thereby avoiding inconsistencies in downstream operations such as encryption and compression. Beyond simple type correspondences, we also capture the operational semantics of C types. Since C pointers permit arithmetic (\emph{e.g.}, addition, subtraction), the address-of operator, and array-style subscripting, we implement equivalent traits in Rust to mirror these behaviors. Likewise, to model the implicit decay of C arrays into pointers, we provide an explicit \verb|cast()| method. Together, these mappings reduce the complexity of translating raw C pointers by supplying direct Rust equivalents for a wide range of pointer classes and operations.

2) \textit{Type Conversion}.
We introduce a custom \verb|CastIntoTyped| trait to handle type conversions in Rust, supporting both integer-to-integer and pointer-to-pointer conversions. For safety, pointer-to-integer casts are allowed, but integer-to-pointer casts are strictly disallowed. For instance, the C expression `\verb|a = (uint16_t)| \verb|b|' is translated as `\verb|a = b.cast::<u16>()|', and `\verb|pi = (int*)pu|' becomes `\verb|pi = pu.cast::<Ptr<i32>>()|'. To simulate C's implicit conversion semantics in Rust's type-inferred context, we also provide a typeless \verb|cast()| method. This proves particularly effective during the repair phase, where it substantially increases the success rate of LLM-driven optimization by efficiently resolving type errors stemming from implicit conversions in the original C code.

3) \textit{Macro/Function Mapping}.
We provide Rust counterparts for native C macros such as \verb|NULL| and \verb|LINE|, as well as for common C library functions including \verb|malloc|, \verb|free|, and \verb|memcpy|. These implementations prioritize safe abstractions over direct invocations of unsafe primitives. For example, rather than relying on the unsafe \verb|std::ptr::copy| to implement \verb|memmove|, we employ a safer alternative. By explicitly incorporating these mappings into the prompt, we guide the LLM to apply them consistently, enabling systematic translations such as converting the C function \verb|name()| into the Rust macro \verb|c_name!()|.

\begin{table}[t]
\caption{An Example of Transformation Patterns}
\centering
\small
\par\noindent
{
 \setlength{\fboxsep}{4pt}
 \setlength{\fboxrule}{0.8pt}
 \fcolorbox{black}{white}{%
  \begin{minipage}{\dimexpr\linewidth-2\fboxsep-2\fboxrule\relax}
    \textbf{Pattern Name:} \\
    \hspace*{1em} Variadic Argument Mapping

    \noindent\rule{\linewidth}{0.2pt}\smallskip

    \textbf{Motivation:} \\
    \hspace*{1em} Rust does not support C-style variadic functions (\emph{e.g.}, \texttt{va\_list}).

    \noindent\rule{\linewidth}{0.2pt}\smallskip 

    \textbf{Consequence:} \\
    \hspace*{1em} Current LLMs cannot directly translate C's variadic features.

    \noindent\rule{\linewidth}{0.2pt}\smallskip 

    \textbf{Solution:} \\
    \hspace*{1em} Define a \texttt{VaList} type and \texttt{va\_format!} macro:\\
    \hspace*{1em} \scriptsize{\texttt{pub type VaList\textless'a\textgreater\ = \&'a [\&'a dyn Display];}}\\
    \small\hspace*{1em} \scriptsize{\texttt{macro\_rules! va\_format \{ \dots\ \} }}\small

    \noindent\rule{\linewidth}{0.2pt}\smallskip 

    \textbf{Example:} \smallskip\\
    C Code:\\
    \scriptsize
    \hspace*{1em} \texttt{int snprintf(char *str, size\_t size, const char *format, \dots) }\\
    \hspace*{1em} \texttt{\{ \dots\ \}}
    \smallskip\\
    \small
    Rust Code:\\
    \scriptsize
    \hspace*{1em}\texttt{pub fn snprintf(mut buf: Ptr\textless u8\textgreater, size: usize, format: Ptr\textless u8\textgreater,}\\
    \hspace*{1em}\texttt{ va: VaList) -> i32 \{}\\
    \hspace*{2em} \texttt{let mut fmt = format.to\_string(); \dots }\\
    \hspace*{2em} \texttt{return length; \}}
  \end{minipage}%
 }%
}
\label{tab:pattern}
\end{table}

4) \textit{Syntax Structure Mapping}.
Certain C control flow constructs lack direct semantic equivalents in Rust, posing a risk of logical errors if translated literally. Specifically, Rust's \verb|for| loop operates differently from C's, C's \verb|do while| has no native counterpart, and \verb|switch| statements are not fully aligned with Rust's \verb|match|. To ensure semantic fidelity, we implement dedicated macros—\verb|c_for|, \verb|c_do|, and \verb|c_switch|—that replicate the exact behavior of their C counterparts. These are introduced to the LLM via targeted prompts, ensuring the original syntax is replaced with the appropriate macro calls.

5) \textit{Operator Mapping}.
We provide Rust equivalents for C-specific operators such as \verb|++|, \verb|--|, \verb|&|, and \verb|sizeof|. The \verb|sizeof| operator is mapped to two distinct macros: \verb|c_sizeof!()| and \verb|c_sizeofval!()|. The increment and decrement operators (\verb|++|, \verb|--|) are implemented as four generic functions to handle both prefix and postfix forms for all integer and pointer types. Notably, the address-of operator \verb|&| is not translated to Rust's reference operators (\verb|&| or \verb|&mut|), but to a dedicated \verb|c_ref!()| macro that correctly obtains a pointer to the value.

6) \textit{Global Variable Mapping}.
Because mutable global variables (\verb|static mut|) are inherently unsafe in Rust, we introduce a \verb|Global<T>| wrapper type to manage them safely. This wrapper employs an internal \verb|Mutex| to ensure thread safety, thereby avoiding the need for unsafe code blocks. To further simplify usage, we provide a \verb|global!()| macro that supports lazy initialization of global variables.

7) \textit{Variadic Argument Mapping}.
Since variadic arguments in C are most commonly used for logging and string formatting, we model the \verb|VaList| type in Rust as a \verb|Slice| of references to values implementing the \verb|Display| trait. To support formatting operations, we provide a \verb|va_format!()| macro, which in turn enables Rust implementations of variadic C functions such as \verb|snprintf|.

These feature mappings are formulated as transformation patterns, each specifying the motivation, consequence, solution, and corresponding code examples, as illustrated in Table \ref{tab:pattern}.
By distilling expertise from senior practitioners, these patterns encode tacit professional knowledge into explicitly reusable C-to-Rust migration guidelines.

To steer the LLM toward accurate and idiomatic translations, we retrieve the most relevant patterns based on the input C code. Specifically, we encode both the code snippet and pattern examples into dense vectors, compute cosine similarity, and select the top-\textit{K} matching patterns. These retrieved patterns are then injected into the LLM prompt using the following structured template.


\smallskip\noindent
{
 \definecolor{acmgray}{RGB}{60,60,60} 
 \setlength{\fboxsep}{0pt} 
 \setlength{\fboxrule}{0.8pt} 
 \fcolorbox{acmgray}{white}{%
  \begin{minipage}{\dimexpr\linewidth-2\fboxrule-15pt\relax}
    \colorbox{acmgray}{%
      \begin{minipage}{\linewidth} 
         \smallskip
         \hspace{4pt}
         \small\textbf{\textcolor{white}{Prompt Template for Translating C Modules to Rust}}
         \smallskip
      \end{minipage}%
    }%
    
    \smallskip
    \hspace{4pt}
    \begin{minipage}{\dimexpr\linewidth-8pt\relax} 
      \small
      Translate the C [\texttt{macro/definition/function}] to Rust.\\
      Patterns: \{\textit{retrieved transformation patterns}\}\\
      Demonstrations: \{\textit{paired C-to-Rust examples}\}\\
      C Source Code: \{\textit{input source code}\}
    \end{minipage}%
    \smallskip
  \end{minipage}%
 }%
}\par

\subsection{Post-Generation Repair}
\label{sec:repair}

Finally, \textsc{EvoC2Rust} refines the translated Rust code through a compilation-driven, cascading repair process that combines syntactic rules with LLM-based correction. For each problematic snippet, it generates multiple repair candidates based on compiler feedback, retaining only those that reduce error counts until no further improvements are possible.

While rule-based techniques efficiently resolve trivial and well-defined errors, LLMs excel at addressing complex semantic issues. A hybrid strategy—applying rules first, followed by LLM refinement—yields the best results. We observe, however, that persistent low-level syntax errors such as bracket mismatches significantly impede both repair methods, particularly rule-based correction. To mitigate this, our approach prioritizes bracket repair before other fixes, ensuring that basic syntax errors are resolved prior to semantic analysis. This allows the Rust compiler to provide clearer diagnostic information for subsequent repair stages.

Specifically, \textsc{EvoC2Rust} implements a three-step repair chain:

1) \textit{Bracket Repair}: An LLM analyzes compiler outputs and error messages to correct mismatched brackets and similar syntax issues. The prompt includes instructions, exemplar fault-correction pairs, the erroneous Rust snippet, and corresponding compilation errors.


\smallskip
\par\noindent
{
 \definecolor{acmgray}{RGB}{60,60,60} 
 \setlength{\fboxsep}{0pt} 
 \setlength{\fboxrule}{0.8pt} 
 \fcolorbox{acmgray}{white}{%
  \begin{minipage}{\dimexpr\linewidth-2\fboxrule-15pt\relax}
    \colorbox{acmgray}{%
      \begin{minipage}{\linewidth} 
         \smallskip
         \hspace{4pt}
         \small\textbf{\textcolor{white}{Prompt Template for Bracket Repair}}
         \smallskip
      \end{minipage}%
    }%
    
    \smallskip
    \hspace{4pt}
    \begin{minipage}{\dimexpr\linewidth-8pt\relax} 
      \small
      Fix the compilation bugs in the following Rust code with the provided compilation error messages, possibly because of mismatched parentheses. Only correct lines that have unmatched parentheses bugs; do not modify any other code.\\
      Demonstrations: \{\textit{example pairs of incorrect/correct Rust code}\} \\
      Rust Source Code: \{\textit{input source code}\} \\
      Compilation Error Messages: \{\textit{input error messages}\}
    \end{minipage}%
    \smallskip
  \end{minipage}%
 }%
}
\smallskip

\begin{table*}[t]
\caption{Statistics of the Datasets}
\label{tab:datasets}
\begin{center}
{
\setlength{\aboverulesep}{0.5pt}
\setlength{\belowrulesep}{0.5pt}
\begin{tabular}{llcccccc}
\toprule
\bf Dataset & \bf Projects &\bf Files &\bf LoC & \bf Functions & \bf Macros (Functions) & \bf Definitions & \bf Test Cases\\
\midrule
Vivo-Bench & 19 projects & 38 & 80$\sim$917 & 200 & 29 & 95 & 113 \\ 
\midrule
\multirow{7}{*}{C2R-Bench} 
 & avl & 9 & 836 & 29 & 29 & 9 & 121\\
 & bzp  & 18 & 2258 & 92 & 82 & 14 & 17\\
 & md5  & 2 & 324 & 8 & 20 & 1 & 37 \\
 & sha256  & 2 & 280 & 9 & 11 & 2 & 12\\
 & rapidlz  & 7 & 748 & 28 & 38 & 10 & 12\\
 & cmptlz  & 25 & 3724 & 122 & 166 & 27 & 23\\
 \cmidrule(lr){2-8}
 & \bf\textit{Total} & 63 & 8170 & 288 & 346 & 63 & 222\\
\bottomrule
\end{tabular}
}
\end{center}
\end{table*}

2) \textit{Rule-Based Repair}: 
We define a set of syntax repair rules as regex-based transformation patterns to address common syntax-level issues. These include adjusting misplaced \verb|derive| macros, removing redundant \verb|cast()| operations, and resolving problematic array access patterns such as \verb|s[s.i]|, which can cause simultaneous mutable and immutable borrow conflicts.

3) \textit{LLM Refinement}: 
In the final step, the LLM addresses residual semantic discrepancies and intricate compilation errors, such as type inconsistencies, unidiomatic usage, or structural mismatches. The prompt incorporates guidelines, fault-correction examples, the erroneous Rust translation, and relevant compilation feedback. These exemplar pairs, which are manually crafted by Rust experts and refined through iterative experimentation, cover common recurring errors to enable the LLM to generalize effective correction strategies (see our repository for detailed examples).


\smallskip
\par\noindent
{
 \definecolor{acmgray}{RGB}{60,60,60} 
 \setlength{\fboxsep}{0pt} 
 \setlength{\fboxrule}{0.8pt} 
 \fcolorbox{acmgray}{white}{%
  \begin{minipage}{\dimexpr\linewidth-2\fboxrule-15pt\relax}
    \colorbox{acmgray}{%
      \begin{minipage}{\linewidth} 
         \smallskip
         \hspace{4pt}
         \small\textbf{\textcolor{white}{Prompt Template for LLM Refinement}}
         \smallskip
      \end{minipage}%
    }%
    
    \smallskip
    \hspace{4pt}
    \begin{minipage}{\dimexpr\linewidth-8pt\relax} 
      \small
      Fix the compilation bugs in the following Rust code according to the compilation information.\\
      Instruction: \{\textit{refinement guidelines}\}\\
      Demonstrations: \{\textit{example pairs of incorrect/correct Rust code}\}\\
      Rust Source Code: \{\textit{input source code}\}\\
      Compilation Information:  \{\textit{input compilation feedback}\}
    \end{minipage}%
    \smallskip
  \end{minipage}%
 }%
}


\section{Experimental Setup}
\label{sec:setup}

We conduct experiments to evaluate the effectiveness of \textsc{EvoC2Rust}, aiming to answer the following research questions.

\begin{itemize}[leftmargin=20pt]
\item \textbf{RQ1}: How effectively does our method translate complete C projects to equivalent safe Rust code? 

\item \textbf{RQ2}: What is the module conversion accuracy achieved by our method?

\item \textbf{RQ3}: To what extent do key components contribute to the overall performance of our method?

\item \textbf{RQ4}: How does our method scale to large projects in terms of accuracy, safety, and efficiency?

\end{itemize}

\subsection{Comparison Methods}
\label{sec:baselines}

We compare \textsc{EvoC2Rust} against three categories of C-to-Rust translation methods: rule-based (C2Rust), LLM-based (Self-Repair and Tymcrat), and hybrid techniques (C2SaferRust). 
We also incorporate a direct LLM prompting baseline that translates code without repository context.
Specifically, the baseline methods evaluated in our study are as follows:

\begin{itemize}[topsep=3pt, leftmargin=20pt]
    \item \textbf{C2Rust}~\cite{c2rust}: a rule-based C-to-Rust translator that converts C code to Rust via AST analysis and manually defined transformation rules.
    
    \item \textbf{C2SaferRust}~\cite{C2SaferRust}: a hybrid technique built on C2Rust, augmented with LLM-based post-processing and test-driven optimization to enhance the safety and idiomaticity of the translated Rust code. We follow the original configuration and use 5 optimization iterations.  
    
    \item \textbf{Self-Repair}~\cite{CRUST-Bench}: a multi-agent self-repair framework for project-level C-to-Rust translation. It incrementally performs file translation and uses iterative compilation and testing feedback to correct errors automatically. We adopt the default setting of 3 self-repair rounds per task.

    \item \textbf{Tymcrat}~\cite{typemigrating}:  a project-level translation method that improves translation quality through type migration. It generates multiple candidate Rust signatures and refines them iteratively using compiler feedback. We use the default configuration, which generates 4 candidate signatures per function.
    
    \item \textbf{Direct Prompting}: a baseline that applies zero-shot LLM prompting for direct C-to-Rust translation. To address its limitations in preserving project-wide consistency and resolving dependencies, we integrate it within our project skeleton framework.
      
\end{itemize}


\subsection{Datasets}
\label{sec:datasets}

We evaluate \textsc{EvoC2Rust} using two benchmarks: the open-source Vivo-Bench~\cite{vivo}, and our newly introduced C2R-Bench, which comprises six industrial projects.

Vivo-Bench is a collection of 19 algorithmic C projects from the 2025 Vivo C-to-Rust Innovation Competition, each containing 1–3 files (33–630 LoC per file). 
Two senior engineers produced the reference Rust translations with LLM assistance. Subsequently, they validated the translations by adding tests to achieve complete function coverage, resulting in a total of 113 top-level test cases.

To assess translation performance in industrial settings, we constructed C2R-Bench using six production projects from Huawei's software ecosystem. These are single-threaded user-space applications using only standard C libraries, characterized by complex multi-file architectures (280–3,724 LoC per file) and extensive cross-file dependencies. Three senior engineers produced the reference Rust implementations via LLM-assisted translation, which were validated using 222 top-level test cases.

Each project in both Vivo-Bench~\cite{vivo} and C2R-Bench provides four artifacts: the original C source code, C test cases, the reference Rust translation, and corresponding Rust test cases. Table~\ref{tab:datasets} summarizes the dataset statistics.

\subsection{Evaluation Strategy and Metrics}
\label{sec:metrics}

Our evaluation assesses C-to-Rust translation performance at two granularities: project-level and module-level, each employing a tailored strategy and specific metrics.

\textbf{Project-level Evaluation}. We benchmark the translation of complete software projects under realistic conditions, where reference implementations are unavailable. This strategy employs incremental compilation: we construct a skeleton with placeholder modules and iteratively replace them with translated code, reverting any modules that cause compilation failures. We employ three metrics to quantify project-level performance:

\begin{itemize}[topsep=3pt, leftmargin=20pt]
\item \textbf{Incremental Compilation Pass Rate (ICompRate)}: The proportion of translated modules that compile successfully when integrated step-wise into the project skeleton, measuring syntactic correctness ~\cite{projectc2rust}.

\item \textbf{Line Acceptance Rate (AccRate)}~\cite{0001KLRRSSA22}: This metric evaluates the fidelity of the initial translation by comparing it to the manually corrected version. Precision captures the percentage of correct lines in the initial output, while Recall measures their preservation in the final code.

\item \textbf{Code Safety Rate (SafeRate)}~\cite{CRustS}: The proportion of generated Rust code that is memory-safe, i.e., free of \verb|unsafe| functions or blocks.
\end{itemize}

\textbf{Module-level Evaluation}.
For a finer-grained analysis, we assess translation accuracy at the module level using a fill-in-the-blank strategy. Given a C module and its reference Rust implementation, we remove the latter, translate the C module, and insert the result into the project to check compilation and test outcomes. Two metrics are used for this assessment:

\begin{itemize}[topsep=3pt, leftmargin=20pt]
\item \textbf{Fill-in Compilation Pass Rate (FCompRate)}: The percentage of generated modules that compile correctly upon integration.
\item \textbf{Test Pass Rate (TestRate)}~\cite{CRUST-Bench}: The percentage of generated modules that pass all associated unit tests.
\end{itemize}

\subsection{Implementation Details}
\label{sec:implementation}
\begin{table*}[t!]
\centering
\caption{Performance of Various Methods in Translating Complete C Projects to Rust}
\label{tab:overall-performance}
\setlength{\tabcolsep}{13pt}
\setlength{\aboverulesep}{0.5pt}
\setlength{\belowrulesep}{0.5pt}
\begin{tabular}{llcccl}
\toprule
\multirow{2}{*}{\textbf{Dataset}} & \multirow{2}{*}{\textbf{Method}} & \multirow{2}{*}{\textbf{ICompRate}} & \multicolumn{2}{c}{\textbf{AccRate}} & \multirow{2}{*}{\textbf{SafeRate}} \\
\cmidrule(lr){4-5}
& &  & \textbf{Precision} & \textbf{Recall} &  \\
\midrule
\multirow{13}{*}{Vivo-Bench} & \multicolumn{5}{l}{\cellcolor{gray!20}\textit{Rule-based and Hybrid Tools}} \\
\cmidrule(lr){2-6}
  & C2Rust~\cite{c2rust}    & \textbf{100} & \textbf{100} & \textbf{100} & 0 \\
  & C2SaferRust~\cite{C2SaferRust} & \textbf{100} & \textbf{100} & \textbf{100} & 60.00 \\
\cmidrule(lr){2-6} & \multicolumn{5}{l}{\cellcolor{gray!20}\textit{LLM-based Methods (DeepSeek-V3)}} \\
\cmidrule(lr){2-6}  
  &Self-Repair~\cite{CRUST-Bench}  & 84.21 & 92.08 & 89.24 & 87.60 \\
  &Tymcrat~\cite{typemigrating}  & \underline{87.35} & 92.11 & 92.44 & 97.88 \\  
  &Direct Prompting  & 35.49 & 55.15 & 51.95 & 77.09 \\  
  &\textsc{EvoC2Rust} (Ours)  & \textbf{100} & \underline{99.83} & \underline{99.86} & \underline{98.00} \\    
\cmidrule(lr){2-6} & \multicolumn{5}{l}{\cellcolor{gray!20}\textit{LLM-based Methods (Qwen3-32B)}} \\
\cmidrule(lr){2-6}  
  &Self-Repair~\cite{CRUST-Bench}  & 52.63 & 33.12 & 38.36 & 77.57 \\
  &Tymcrat~\cite{typemigrating}  & 81.76 & 76.82 & 74.47 & 87.11 \\  
  &Direct Prompting  & 26.23 & 21.72 & 27.51 & 89.81 \\  
  &\textsc{EvoC2Rust} (Ours)  & 87.65 & 87.92 & 83.45 & \textbf{98.22} \\ 
\midrule  
\multirow{13}{*}{C2R-Bench} & \multicolumn{5}{l}{\cellcolor{gray!20}\textit{Rule-based and Hybrid Tools}} \\
\cmidrule(lr){2-6}
  & C2Rust~\cite{c2rust}  & \textbf{99.28} & \textbf{98.99} & \textbf{98.98} & 1.83 \\
  & C2SaferRust~\cite{C2SaferRust} & \textbf{99.28} & 97.47 & \underline{97.68} & 48.24 \\
\cmidrule(lr){2-6} & \multicolumn{5}{l}{\cellcolor{gray!20}\textit{LLM-based Methods (DeepSeek-V3)}} \\
\cmidrule(lr){2-6}  
  &Self-Repair~\cite{CRUST-Bench}  & 49.21 & 72.35 & 13.79  & 83.92 \\
  &Tymcrat~\cite{typemigrating}  & 72.02 & 77.82  & 74.94 & 95.92 \\  
  &Direct Prompting  & 56.67 & 55.65 & 52.70 & 82.43 \\  
  &\textsc{EvoC2Rust} (Ours)  & \underline{93.84} & \underline{97.56} & 97.34 & \textbf{97.41} \\    
\cmidrule(lr){2-6} & \multicolumn{5}{l}{\cellcolor{gray!20}\textit{LLM-based Methods (Qwen3-32B)}} \\
\cmidrule(lr){2-6}  
  &Self-Repair~\cite{CRUST-Bench}  & 39.40 & 45.38 & 10.65 & 87.15 \\
  &Tymcrat~\cite{typemigrating}  & 70.56 & 66.72 & 65.91 & 82.75 \\  
  &Direct Prompting  & 35.29 & 24.01 & 30.41 & 93.63 \\  
  &\textsc{EvoC2Rust} (Ours)  & 82.35 & 85.75 & 82.59 & \underline{97.40} \\ 
\bottomrule
\end{tabular}
\begin{minipage}{0.95\linewidth}
\footnotesize
\hspace*{5em}* \textbf{Bold} and \underline{underline} denote the best and second-best results, respectively.
\end{minipage}
\end{table*}
\begin{table}[t!]
\centering
\caption{Effectiveness of \textsc{EvoC2Rust} in Module-Level C-to-Rust Translation}
\label{tab:function-translation}
\setlength{\tabcolsep}{10pt}
\setlength{\aboverulesep}{0.5pt}
\setlength{\belowrulesep}{0.5pt}
\begin{tabular}{llcc}
\toprule
\textbf{Dataset} & \textbf{Project} & \textbf{FCompRate} & \textbf{TestRate}  \\
\midrule
\rowcolor[HTML]{EFEFEF} 
\multicolumn{4}{c}{\textbf{DeepSeek-V3}} \\
\hline
Vivo-Bench  & 19 projects & 99.07 & 98.50 \\
\cmidrule(lr){1-4}
\multirow{7}{*}{C2R-Bench} & avl  & 100.00 & 92.53 \\
  & bzp  & 95.21 & 92.55 \\
  & md5 & 96.55 & 86.21 \\
  & sha256 & 100.00 & 100.00 \\
  & rapidlz & 96.05 & 92.11 \\
  & cmptlz & 86.98 & 86.35 \\
\cmidrule(lr){2-4}
  & \textbf{Subtotal} & \textbf{92.25} & \textbf{89.53}  \\
\midrule
\rowcolor[HTML]{EFEFEF}
\multicolumn{4}{c}{\textbf{Qwen3-32B}} \\
\hline
Vivo-Bench  & 19 projects & 87.65 & 84.57 \\
\cmidrule(lr){1-4}
\multirow{7}{*}{C2R-Bench} & avl  & 97.01 & 82.09 \\
  & bzp  & 89.36 & 88.83 \\
  & md5 & 89.66 & 89.21 \\
  & sha256 & 90.91 & 90.91 \\
  & rapidlz & 78.95 & 77.63 \\
  & cmptlz & 70.79 & 68.89 \\
\cmidrule(lr){2-4}
  & \textbf{Subtotal} & \textbf{80.63} & \textbf{77.91}  \\
  \bottomrule
\end{tabular}
\end{table}

We implement \textsc{EvoC2Rust} using DeepSeek-V3~\cite{deepseekv3} and Qwen3-32B~\cite{Qwen3-32B} as the foundation models. 
Translation tasks use greedy decoding with \texttt{max\_tokens} set to 4096. We leverage Tree-sitter v0.22.3~\cite{treesitter} for C code parsing, the standard rustc~\cite{rustc} for compilation, and the BGE-M3 model~\cite{BGE-M3} to generate embeddings and retrieve the top-10 most relevant patterns based on cosine similarity. Iteration limits are enforced to ensure practicality: 5 rounds for bracket repair and 3 rounds for LLM refinement. 

We established verified ground-truth implementations to evaluate the line acceptance rate metric. For each project, we first translated the code using the target methods. The resulting Rust artifacts were then iteratively compiled and tested. Any compilation or test failures were manually repaired by three senior engineers with assistance from Claude Sonnet 4~\cite{claude4} until all tests passed.


\section{Results and Analysis}
\label{sec:results}

\subsection{Main Results (RQ1)}
\label{sec:RQ1}

\textsc{EvoC2Rust} demonstrates superior performance in project-level C-to-Rust translation, as summarized in Table~\ref{tab:overall-performance}. It consistently outperforms all baseline methods across both datasets, achieving the highest overall performance in terms of compilation success, line acceptance, and code safety.

\textsc{EvoC2Rust} demonstrates a significant advantage in generating memory-safe Rust code compared to rule-based and hybrid baselines. Although C2Rust and C2SaferRust achieve near-perfect compilation pass rates (\emph{e.g.}, 100\% on Vivo-Bench), their safety performance is considerably lower. C2Rust exhibits code safety rates of only 0\% and 1.83\% on the two datasets, while C2SaferRust, despite LLM-enhanced optimization, reaches merely 60\% and 48.24\%. This confirms that syntactic translation, even with LLM post-processing, fails to reliably meet Rust's safety requirements. In contrast, our method proactively prevents unsafe code by leveraging safety-guaranteed feature mappings.

Among LLM-based approaches, \textsc{EvoC2Rust} demonstrates consistent superiority. On Vivo-Bench, it surpasses all baselines by 12.65\% to 64.51\% in compilation success and 7.57\% to 46.30\% in line acceptance using DeepSeek-V3, while maintaining a 98\% code safety rate. The performance gap widens on the complex C2R-Bench, where it achieves 93.84\% compilation success (21.82\%-44.63\% higher than baselines) and over 97\% line acceptance. This trend is also observed with the Qwen3-32B model, highlighting the effectiveness of our skeleton-guided, feature mapping-augmented translation paradigm.

Specifically, \textsc{EvoC2Rust} improves compilation success by 44.63\% and line acceptance by 54.38\% over Self-Repair on C2R-Bench with DeepSeek-V3 (42.95\%/55.66\% with Qwen3-32B). Self-Repair's file-level processing often exceeds LLM context limits, leading to code truncation. 
Compared to Tymcrat, \textsc{EvoC2Rust} achieves a 21.82\% higher compilation rate on C2R-Bench with DeepSeek-V3 (11.79\% with Qwen3-32B). While Tymcrat employs diverse signature generation and context enrichment to reduce initial errors, its lack of expert-derived structured translation guidance ultimately constrains its effectiveness in complex scenarios. 
Direct Prompting performs poorest, with a 38.42\% average compilation success and low line acceptance (39.13\% precision, 40.64\% recall). This is primarily due to the inherent linguistic gaps between C and Rust. 


\smallskip\noindent
{
 \setlength{\fboxsep}{5pt}
 \setlength{\fboxrule}{0.8pt}
 \fcolorbox{black}{gray!10}{
  \begin{minipage}{\dimexpr0.43\textwidth\fboxsep-2\fboxrule\relax}
   \textbf{Answer to RQ1.} \textsc{EvoC2Rust} surpasses all baseline methods in project-level C-to-Rust translation, demonstrating the highest overall performance across compilation success, line acceptance, and code safety on both open-source and industrial benchmarks.
  \end{minipage}%
 }%
}\par

\begin{table*}[t!]
\centering
\caption{Ablation Results on Key Components of \textsc{EvoC2Rust}}
\label{tab:component-ablation}
\setlength{\tabcolsep}{13pt}
\begin{tabular}{lcccccc}
\toprule
\multirow{2}{*}{\textbf{Variant}} & \multicolumn{2}{c}{\textbf{Project Translation}} & \multicolumn{2}{c}{\textbf{Module Translation}} \\
\cmidrule(lr){2-3}
\cmidrule(lr){4-5}
 & \textbf{ICompRate} & \textbf{SafeRate} & \textbf{FCompRate} & \textbf{TestRate} \\
\midrule
\textsc{EvoC2Rust} (Ours) & \textbf{93.84} & 97.41 & \textbf{92.25} & \textbf{89.53} \\
-w/o Repair\#3 & 89.57 (\textcolor[HTML]{FFA9A9}{\ -4.27}) & 97.40 (\textcolor[HTML]{FFA9A9}{\ -0.01}) & 87.23
(\textcolor[HTML]{FFA9A9}{\ -5.02}) & 85.37 (\textcolor[HTML]{FFA9A9}{\ -4.16}) \\
-w/o Repair\#2-3 & 75.24 (\textcolor[HTML]{FF6C6C}{-18.60}) & \textbf{97.46} (\textcolor[HTML]{A0A0A0}{\ +0.05}) & 82.21 (\textcolor[HTML]{FF6C6C}{-10.04}) & 80.92 (\textcolor[HTML]{FF6C6C}{\ -8.61})\\
-w/o Repair\#1-3 & 74.29 (\textcolor[HTML]{FF5959}{-19.55}) & 97.21 (\textcolor[HTML]{FFA9A9}{\ -0.20}) & 81.06 (\textcolor[HTML]{FF5959}{-11.19}) & 79.91 (\textcolor[HTML]{FF5959}{\ -9.62})\\
-w/o Feature Mapping and Repair\#1-3 & 56.67 (\textcolor[HTML]{FF0000}{-37.17}) & 82.43 (\textcolor[HTML]{FF6C6C}{-14.98}) & 33.00
 (\textcolor[HTML]{FF0000}{-59.25}) & 30.27
 (\textcolor[HTML]{FF0000}{-59.26})  \\
\bottomrule
\end{tabular}
\begin{minipage}{0.95\linewidth}
\footnotesize
\hspace*{2em}* Repair\#1: bracket repair; Repair\#2: rule-based repair; Repair\#3: LLM refinement.
\end{minipage}
\end{table*}

\subsection{Module Translation (RQ2)}
\label{sec:RQ2}

We further evaluate the module translation accuracy of \textsc{EvoC2Rust}, which serves as its core technical component. As shown in Table~\ref{tab:function-translation}, \textsc{EvoC2Rust} exhibits strong performance at the module level. Using DeepSeek-V3, it achieves 100\% compilation and 99.07\% test pass rates on Vivo-Bench, while maintaining robust results on C2R-Bench (92.25\% compilation, 89.53\% test pass). Even with the smaller Qwen3-32B model which has 20× fewer parameters, \textsc{EvoC2Rust} still delivers competitive performance (80.63\% compilation, 77.91\% test pass), demonstrating its model-agnostic applicability.

A notable observation is the strong correlation between test pass rates and compilation success: modules that compile correctly generally also pass functional tests. This consistency results from our method's dual mechanisms. Feature mapping prevents semantic errors, and compiler-guided repair maintains logically sound error correction.

Our analysis confirms that project complexity significantly impacts translation quality. Complex projects with intricate control flows and data dependencies consistently challenge LLM comprehension capabilities.  For instance, the \verb|cmptlz| project, featuring complex macros and dependencies, achieves test pass rates of 86.35\% and 70.79\% across models. In contrast, the simpler \verb|sha256| project attains optimal 100\% pass rates.


\smallskip\noindent
{
 \setlength{\fboxsep}{5pt}
 \setlength{\fboxrule}{0.8pt}
 \fcolorbox{black}{gray!10}{
  \begin{minipage}{\dimexpr0.43\textwidth\fboxsep-2\fboxrule\relax}
   \textbf{Answer to RQ2.} Our method demonstrates robust performance in module-level translation, achieving 92.25\%–100\% compilation success and 77.91\%–99.07\% test pass rates across model scales, confirming its effectiveness in handling projects of varying complexity.
  \end{minipage}%
 }%
}\par

\subsection{Ablation Study (RQ3)}
\label{sec:RQ3}

To assess the contribution of each component in \textsc{EvoC2Rust} and validate our design choices, we perform an ablation study by incrementally removing its key components: feature mapping, bracket repair, rule-based repair, and LLM refinement. Each variant is evaluated on C2R-Bench benchmark using the DeepSeek-V3 model. 
In our cascaded repair chain, later repairs depend on the output of earlier steps. Removing an early component while keeping later ones would thus propagate unresolved errors, preventing compilation or testing and invalidating the assessment of downstream components. To avoid this issue and ensure a meaningful evaluation, we remove components in reverse order, beginning with the final step (LLM refinement).

As illustrated in Table~\ref{tab:component-ablation}, every component in \textsc{EvoC2Rust} contributes critically to its overall effectiveness.
Among them, the feature mapping mechanism is the most impactful component. By ablating this mechanism, both ICompRate and FCompRate drop sharply from 74.29\% to 56.67\% and from 81.06\% to 33\%, respectively. The impact on semantic correctness is also substantial, with TestRate decreasing from 79.91\% to 30.27\%. This result highlights the mechanism's essential role in bridging C and Rust constructs through predefined transformation patterns that map C idioms to safe Rust equivalents where direct counterparts are absent.

The cascading repair chain also facilitates C-to-Rust translation. Removing all three repairs (-w/o repair\#1-3) results in considerable performance degradation, with ICompRate dropping by 19.55\%, FCompRate by 11.19\%, and TestRate by 9.62\%. Within this chain, rule-based repair (repair\#2) contributes most significantly, raising the compilation pass rate by 9.68\% and the test pass rate by 4.45\% on average. LLM refinement (repair\#3) further enhances syntactic correctness by 4.27\% and semantic correctness by 8.98\%, without compromising code safety.

Code safety remains consistently high (around 97\%) across nearly all ablation settings. This stability is attributable to the feature mapping mechanism’s encapsulation of low-level operations within verified safe constructs. Only when both feature mapping and repairs are disabled does safety drop sharply to 82.43\%, demonstrating the necessity of our integrated design.


\smallskip\noindent
{
 \setlength{\fboxsep}{5pt}
 \setlength{\fboxrule}{0.8pt}
 \fcolorbox{black}{gray!10}{
  \begin{minipage}{\dimexpr0.43\textwidth\fboxsep-2\fboxrule\relax}
   \textbf{Answer to RQ3.} Each component in \textsc{EvoC2Rust} plays a crucial role in its overall performance. The feature mapping mechanism provides a safe and reliable basis by bridging C-to-Rust semantic gaps, while the cascading repair chain fixes remaining errors to maximize syntactic and semantic correctness.
  \end{minipage}%
 }%
}\par

\subsection{Scalability on Larger Projects (RQ4)}
\label{sec:RQ4}

To evaluate the scalability and performance of our method on large-scale codebases, we conduct experiments on 10 larger projects (9,973 - 91,588 LoC) from RepoTransBench~\cite{repotransbench}, a benchmark derived from highly-starred GitHub repositories. During module translation, we exclude functions that depend on third-party libraries, as our current implementation primarily supports dependencies on the ISO C standard library. Using DeepSeek-V3 as the backbone model, \textsc{EvoC2Rust} achieves acceptable performance, with an ICompRate of 75.61\% and a SafeRate of 96.20\%. We do not report line acceptance rates due to the prohibitive cost of manually creating verified Rust references for projects at this scale.

By integrating results from all evaluated datasets (\emph{i.e.}, Vivo-Bench, C2R-Bench, and RepoTransBench), we further analyze translation performance across different project scales.

\begin{figure}[t]
\centerline{\includegraphics[width=0.36\textwidth, trim=0 8 0 15, clip]{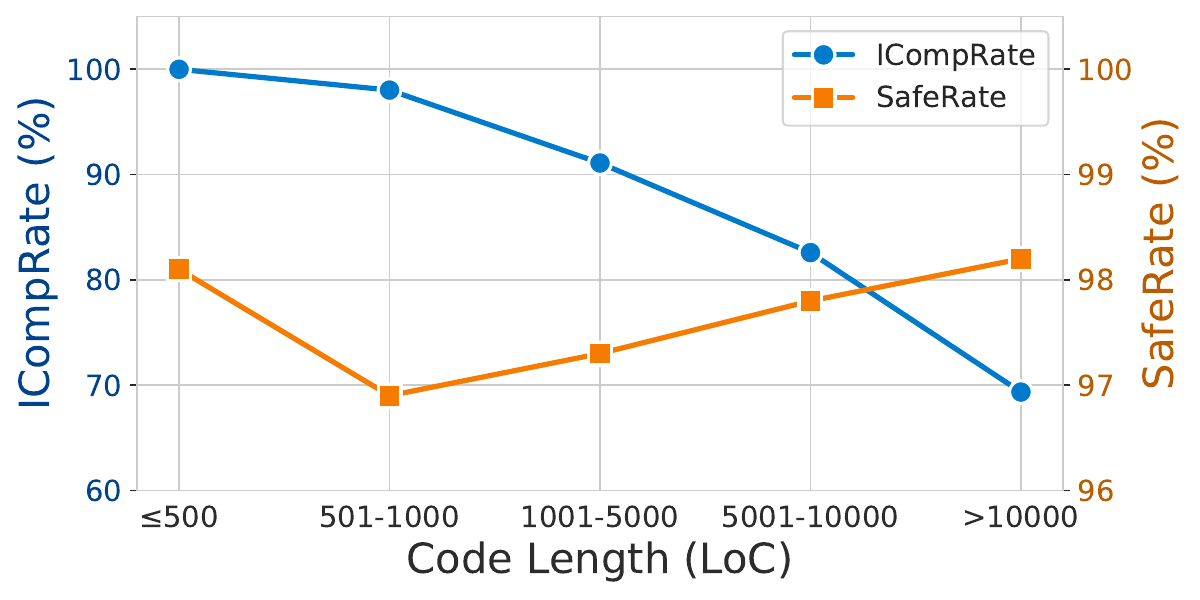}}
    \caption{Translation performance at different project scales.}
    \label{fig:loc-project}
    \Description{The figure shows a line chart comparing translation performance across projects with different code lengths. The blue line represents ICompRate, which decreases steadily from 100\% for projects under 500 lines of code to about 69\% for projects over ten thousand lines. The orange line indicates SafeRate, which starts around 98\%, dips for medium-sized projects, and then gradually increases to about 98\% for the largest projects. The x-axis categorizes projects into five code-length ranges, from under 500 LOC to over ten thousand LOC. Two y-axes are used to separately display the percentage scales for ICompRate and SafeRate.}
\end{figure}

\textbf{Translation Accuracy vs. Project Scale}. As shown in Figure~\ref{fig:loc-project}, compilation success (ICompRate) exhibits a clear negative correlation with project size. While the rate declines gradually from 100\% for small projects ($\le$500 LoC) to approximately 69\% for large projects (>10,000 LoC), it does not collapse. This robustness to scaling is due to our skeleton-guided approach, which effectively manages complex inter-module dependencies.

\textbf{Translation Safety vs. Project Scale}. The SafeRate remains consistently high (around 97\%) across all project sizes, and even exhibits a slight improvement in larger projects. We attribute this stability to our safety-guaranteed feature mappings, which encapsulate low-level operations into a relatively fixed set of unsafe code blocks. As project size increases, the proportion of such unsafe code naturally decreases, resulting in the observed improvement in SafeRate.

\begin{figure}[t]
\centerline{\includegraphics[width=0.36\textwidth, trim=0 8 0 15, clip]{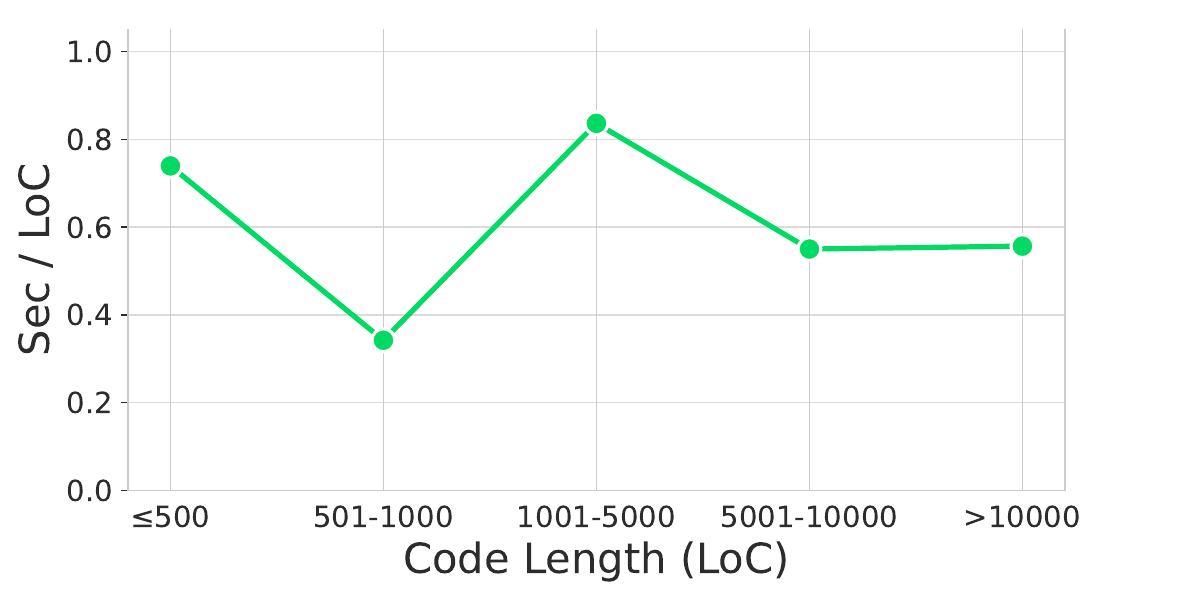}}
    \caption{Translation time at different project scales.}
    \label{fig:loc-time}
    \Description{The figure presents a line chart illustrating translation time per line of code across varying project sizes. The x-axis divides projects into five code-length intervals, ranging from under 500 to over ten thousand lines. The y-axis measures time in seconds per line of code. The trend begins at approximately 0.75 seconds per line for the smallest projects (<500 LOC), declines to around 0.35 seconds for projects between 501–1000 LOC, peaks at about 0.85 seconds for the 1001–5000 LOC range, and subsequently decreases and stabilizes near 0.55 seconds per line for the largest projects.}
\end{figure}

\textbf{Translation Efficiency vs. Project Scale}. Figure~\ref{fig:loc-time} presents the translation efficiency, quantified as seconds per line of code (Sec/LoC), across projects of varying scales. The per-line translation time ranges from approximately 0.35 to 0.85 seconds, averaging 0.63 seconds. Notably, for projects exceeding 5,000 LoC, efficiency stabilizes at approximately 0.55 Sec/LoC, demonstrating strong scalability of our method. These performance fluctuations show no correlation with code length and are instead determined by project-specific characteristics. The primary factor influencing this variation is the number of LLM refinement iterations required, which depends directly on the inherent complexity of the code in each project.


\smallskip\noindent
{
 \setlength{\fboxsep}{5pt}
 \setlength{\fboxrule}{0.8pt}
 \fcolorbox{black}{gray!10}{
  \begin{minipage}{\dimexpr0.43\textwidth\fboxsep-2\fboxrule\relax}
   \textbf{Answer to RQ4.} Our method exhibits strong scalability on large-scale projects, maintaining resilient compilation rates, high safety guarantees, and stable efficiency without performance collapse.
  \end{minipage}%
 }%
}

\subsection{Case Study}

\begin{figure*}[t]
    \centering
    \includegraphics[width=\textwidth]{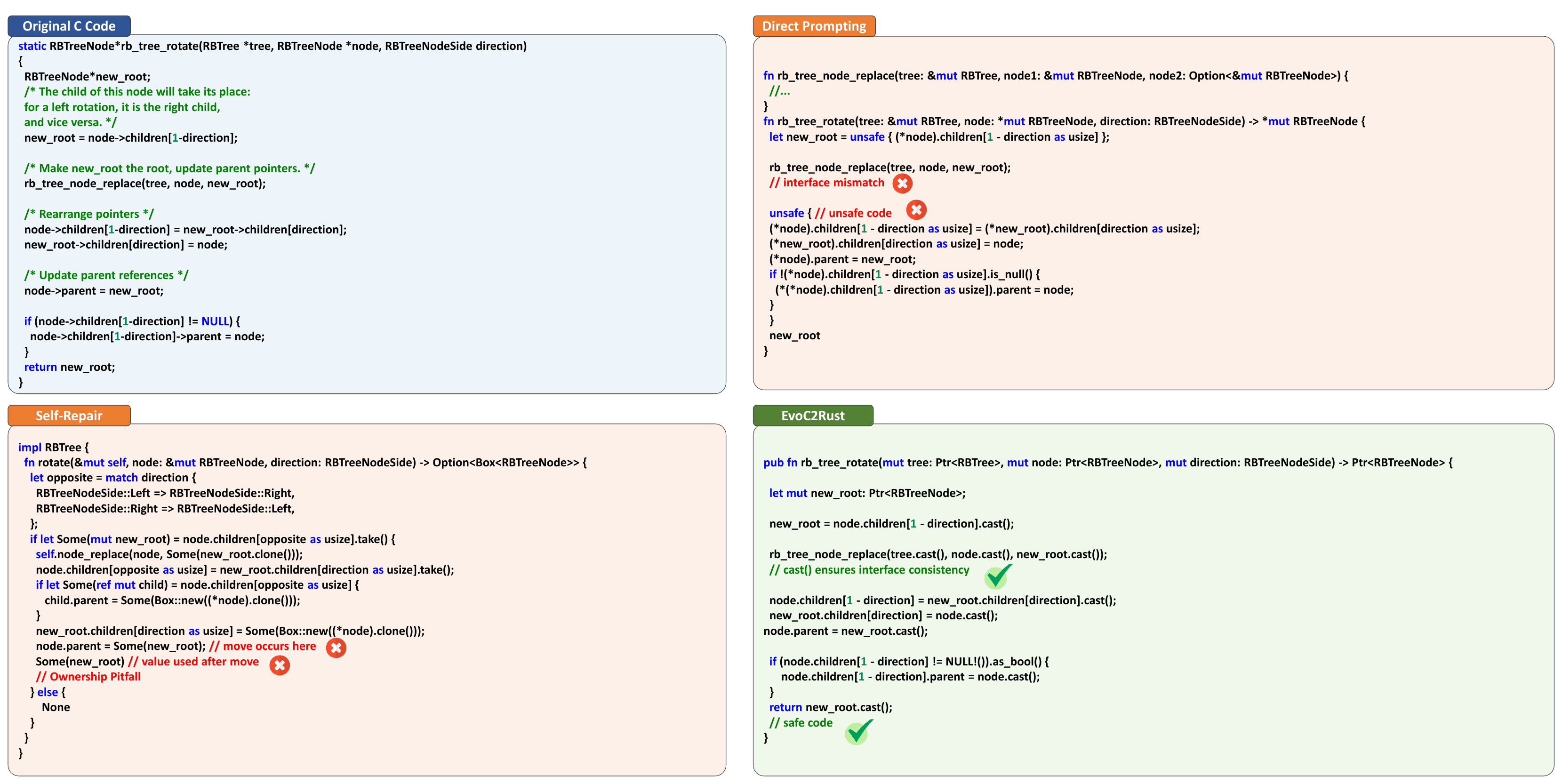}
    \caption{A case study on translating the \texttt{rb\_tree\_rotate} function.}
    \label{fig:case-study}
    \Description{The figure compares four versions of the rb_tree_rotate function: the original C implementation, a direct prompting translation in Rust, a self-repair version, and the EvoC2Rust result. The top-left panel shows the original C code. The top-right panel displays Rust code generated by direct prompting, which includes unsafe blocks and interface mismatches. The bottom-left panel presents Rust code generated by self-repair, which introduces ownership and move errors. The bottom-right panel depicts the EvoC2Rust translation, which uses consistent pointer casting and produces safe, compilable Rust code. Overall, the figure illustrates how EvoC2Rust successfully avoids the unsafe constructs and ownership pitfalls observed in alternative translation approaches.}
\end{figure*}

We present a case study on the \texttt{rb\_tree\_rotate} function, a core operation in a red-black tree implementation, to illustrate the effectiveness of \textsc{EvoC2Rust} (Figure~\ref{fig:case-study}). This function involves intricate pointer manipulation, characteristic of low-level C code that is difficult to translate into safe, idiomatic Rust. Its complexity exposes common limitations of existing translation approaches.

As shown in Figure~\ref{fig:case-study}, baseline methods exhibit critical failures. Direct Prompting produces a project-level interface mismatch by passing a raw pointer (\texttt{*mut}) where a mutable reference (\texttt{\&mut}) is required, revealing a lack of project-wide context awareness. Moreover, its heavy use of raw pointers necessitates unsafe blocks, contradicting Rust's memory safety goals. Self-Repair attempts a more idiomatic translation using \texttt{Box<T>}, but introduces a function-level ownership error (\texttt{use of moved value}), demonstrating the difficulty of refactoring C-style memory management into safe Rust idioms without  safety-guaranteed feature mappings.

In contrast, \textsc{EvoC2Rust} generates a correct and safe translation. This is achieved through our feature mapping mechanism, which employs a unified \texttt{Ptr<T>} smart pointer to preserve C-like pointer semantics while ensuring project-wide type consistency. Combined with a flexible \texttt{.cast()} method for type conversions, this approach resolves both the interface mismatch and ownership issues that cause other methods to fail. By abstracting low-level C operations into a safe Rust layer, \textsc{EvoC2Rust} succeeds where other techniques fall short.

\section{Limitations and Threats to Validity}
We have identified the following limitations and potential threats to the validity of our method:

\textbf{Internal Validity}. The main internal threat comes from correctness verification, which currently relies on predefined test cases. Adopting more robust validation techniques like \textit{fuzzing}~\cite{KadronNPBPS24} and \textit{self-debugging}~\cite{KangCYL25,shi2024code} could enhance test coverage and uncover subtle semantic errors.
For line acceptance rate measurement, we use Claude-generated outputs verified by human experts as ground truth. While this manual validation ensures correctness beyond automated testing, it may not scale to larger datasets. Future work should explore semi-automated verification to balance accuracy and efficiency.

\textbf{External Validity}. Three primary threats affect external validity: 
1) \textit{Generalizability}: While our translation pipeline is designed to support multiple language pairs, the current implementation specifically targets C-to-Rust translation.
Moreover, due to resource constraints, our evaluation is limited to DeepSeek-V3 and Qwen3-32B. These choices do not, however, affect the validity of our framework evaluation, as the primary contribution lies in the architecture itself rather than in model comparison. Extending the evaluation to include additional languages and models remains an important direction for future work.
2) \textit{Data leakage}: Our method uses LLMs to translate C code to Rust. Since these models may have been pre-trained on public benchmarks, we mitigate this risk by including six industrial projects alongside open-source datasets in our evaluation. 
3) \textit{Dataset characteristics}: The current benchmarks are limited to single-threaded, user-level C projects with ISO C standard library dependencies. Future work should consider more complex settings such as multithreading, third-party libraries, and kernel-level code.

\section{Related Work}
\label{sec:related}

\subsection{C-to-Rust Translation}
Existing work for automatic C-to-Rust translation falls into three broad categories: rule-based, LLM-based, and hybrid techniques. 

\textbf{Rule-based Methods}
rely on manually designed rules to transform C code into Rust. The C2Rust transpiler~\cite{c2rust} exemplifies this approach by using Clang's AST to generate Rust code, though the output relies heavily on \texttt{unsafe} blocks. Subsequent work has aimed to improve safety and idiomaticity: Emre et al.~\cite{EmreSDH21, EmreSDH23} used compiler feedback for pointer safety; Zhang et al.~\cite{ZhangDYW23} applied static ownership analysis; Ling et al.~\cite{CRustS} and Hong et al.~\cite{HongR23, HongR24a, HongR24b, Forcrat} focused on API safety and specific constructs; Larson et al.~\cite{Larsen24} proposed pointer derivation graphs to translate unsafe Rust to safer Rust; Wu et al.~\cite{WuD25} retyped polymorphic C void pointers into uses of Rust generics. While effective, these methods require significant manual effort and often produce Rust code that is both unsafe and non-idiomatic.

\textbf{LLM-based Methods}
leverage LLMs to generate idiomatic Rust code without relying on manually defined rules. 
Eniser et al.~\cite{FLOURINE} introduced FLOURINE, which uses fuzz testing to validate semantic equivalence between C and Rust code.
Yang et al.~\cite{vert} adapted MSWasm-based testing for multi-language translation to Rust.
Luo et al.~\cite{IRENE} proposed IRENE, an LLM-based framework that integrates rule-augmented retrieval and structured summarization for function-level translation.
Nitin et al.~\cite{SpecTra} enhanced LLM translation by integrating static specifications and I/O tests.
Xu et al.~\cite{XuH25} optimized type migration by leveraging data flow graphs.
Farrukh et al.~\cite{SafeTrans} developed a multi-agent framework featuring iterative error correction. 
However, LLM-based translation suffers from lower accuracy due to the lack of parallel C-Rust training data~\cite{RustRepoTrans}.

\textbf{Hybrid Methods}
seek to combine the strengths of both paradigms. For instance, C2SaferRust~\cite{C2SaferRust} and PR$^2$~\cite{PR2} employ LLMs to refine rule-based translations, improving safety and idiomaticity while preserving functional equivalence through testing. 
SACTOR~\cite{SACTOR} delegates different tasks to each paradigm, applying C2Rust for data types and an LLM for global variables and functions.
Our method extends this direction with a two-stage framework: it first uses linguistic feature mappings (\emph{i.e.}, transformation rules) to guide LLM translation, then applies a compilation-driven repair mechanism combining LLM refinement with static analysis. It achieves an effective balance between translation accuracy and safety in automated C-to-Rust migration.

\subsection{Project-level Code Translation with LLMs}

While most C-to-Rust translation methods target small code units, recent work explores LLM-based project-level translation. These approaches typically decompose a C project into dependency-ordered units, translate them sequentially, and reassemble the results.

Shiraishi et al.~\cite{projectc2rust} pioneered this direction by using project metadata to maintain consistency, though their method prioritizes compilation success over functional correctness. Syzygy~\cite{Syzygy} and RustMap~\cite{RustMap} employ a strategy of translating both functions and test cases while utilizing feedback mechanisms for repair. Nevertheless, these approaches necessitate complete dependency contexts (occasionally through manual intervention) and exhibit limited scalability when applied to complex industrial codebases.
Khatry et al.~\cite{CRUST-Bench} and Ou et al.~\cite{RustRepoTrans} introduced repository-scale benchmarks and self-repair techniques, yet rely on annotated Rust interfaces rarely available in practice. Hong et al.~\cite{typemigrating} addressed this with Tymcrat, a type inference-based method for automating signature translation.

Differing from existing approaches that translate modules sequentially following dependency order and require full project contexts, \textsc{EvoC2Rust} introduces a skeleton-guided strategy, which first constructs a compilable Rust skeleton and then executes incremental function translation. 
Beyond enabling concurrent processing, this design also enhances correctness by isolating the LLM from cross-module dependencies.

\section{Conclusion}
\label{sec:conclusion}

We present \textsc{EvoC2Rust}, a novel LLM-powered framework for automated translation of complete C projects to Rust. 
By leveraging feature mapping-enhanced LLMs, \textsc{EvoC2Rust} performs skeleton-guided code translation, augmented with hybrid error correction that combines LLM capabilities with static analysis. Extensive evaluation on both open-source and industrial benchmarks demonstrates that \textsc{EvoC2Rust} consistently achieves superior overall performance in syntax accuracy, semantic equivalence, and memory safety. 
Future work will extend the approach to more complex settings, including multithreading, third-party libraries, and kernel-level code across diverse application domains.


\begin{acks}
This research is funded by the National Key Research and Development Program of China (Grant No. 2023YFB4503802), the National Natural Science Foundation of China (Grant No. 62232003), and the Natural Science Foundation of Shanghai (Grant No. 25ZR1401175).
\end{acks}

\balance
\bibliographystyle{ACM-Reference-Format}
\bibliography{ref}
\clearpage
\end{document}